\documentclass[10pt,a4paper,twoside]{article}
\usepackage{epsfig}
\usepackage{baltlat6}
\usepackage{array}
\usepackage{here}
\usepackage{amssymb,amsmath}

\newcommand{\p}{\partial}

\begin{document}
\ \
\vspace{0.5mm}
\setcounter{page}{277}

\titlehead{Baltic Astronomy, vol.\,25, 1--8, 2016}

\titleb{Bar formation in the Milky Way type galaxies}

\begin{authorl}
\authorb{E.V. Polyachenko}{1} and
\authorb{P. Berczik}{2,3,4} and
\authorb{A. Just}{2}
\end{authorl}

\begin{addressl}
\addressb{1}{Institute of Astronomy, Russian Academy of Sciences,\\  Pyatnitskaya 48, Moscow 119017, Russia; epolyach@inasan.ru}
\addressb{2}{Astronomisches Rechen-Institut am ZAH,\\ M\"onchhofstr. 12-14, D-69120 Heidelberg, Germany; just@ari.uni-heidelberg.de}
\addressb{3}{Main Astronomical Observatory, National Academy of Sciences of Ukraine, \\MAO/NASU, 27 Akad. Zabolotnoho St. 03680 Kyiv, Ukraine; berczik@mao.kiev.ua}
\addressb{4}{National Astronomical Observatories of China, Chinese Academy of Sciences, \\NAOC/CAS, 20A Datun Rd., Chaoyang eDistrict Beijing 100012, China}
\end{addressl}

\submitb{Received: 2016 June 24; accepted: 2016 xxx xx}

\begin{summary} 
	Many barred galaxies, possibly including the Milky Way, have cusps in the centres. There is a widespread belief, however, that usual bar instability taking place in bulgeless galaxy models is impossible for the cuspy models, because of the presence of the inner Lindblad resonance for any pattern speed. At the same time there are numerical evidences that the bar instability can form a bar. We analyse this discrepancy, by accurate and diverse N-body simulations and using the calculation of normal modes. We show that bar formation in cuspy galaxies can be explained by taking into account the disc thickness. 
	The exponential growth time is moderate for typical current disc masses (about 250 Myr), but considerably increases (factor 2 or more) upon substitution of the live halo and bulge with a rigid halo/bulge potential; meanwhile pattern speeds remain almost the same. Normal mode analysis with different disc mass favours a young bar hypothesis, according to which the bar instability saturated only recently.  	
\end{summary}


\resthead{Bar formation in galaxies}
{E. V. Polyachenko, P. Berczik, A. Just}

\sectionb{1}{INTRODUCTION}

When studying bar formation in disc galaxies, it is a habit already to use models without bulges. These models show fast bar formation, followed by a growth of the central density and formation of a pseudo-bulge, followed by bar decay and possibly recurrent bar formation. This scenario, however, contradicts to Hubble deep field observations, according to which less bars are seen at high cosmological redshifts, $z \ge 0.5$ (Abraham et al. 1999, Merrifield et al. 2000). Besides the ratio ${\cal R} = R_\textrm{c}/R_\textrm{b}$ of the corotation radius to bar radius (half bar length) is usually in the range between 0.9 and 1.3 (Binney \& Tremaine 2008).  This ratio is nearly 1 just after the bar formation, but it grows eventually due to slowdown and shortening of the bar.

On the other hand, in models with cusps when the density $\rho$ rises as $r^{-\alpha}$, there is a problem with the inner Lindblad resonance (ILR), obeying $m(\Omega(R) - \Omega_\textrm{p}) = \kappa(R)$ for a given pattern speed $\Omega_\textrm{p}$ ($\Omega(R)$ and $\kappa(R)$ standardly denote angular velocity and epicyclic frequency). Since we are interested in bar formation, an azimuthal number $m=2$ is adopted in this paper. According to a theoretical point of view, the ILR damps waves (Mark 1971, 1974) and prevents bar formation (Toomre 1981). In Figure 1a we show the so called {\it precession} curves $\Omega_\textrm{pr} \equiv \Omega - \kappa/2$, which give positions of the ILR, for a cored (or bulgeless) model and for a model with the cusp. Physically, $\Omega_\textrm{pr}(R)$ determines a precession rate of nearly circular orbits and plays an important role in the formation of bars in razor thin discs (Polyachenko 2004). The cored profile has a maximum, so any pattern speed above the maximum is available for bar formation. The cuspy profile does not allow bar formation at any pattern speed. This speculation along with Hubble deep field observations lead Sellwood (2000) to the suggestion that most real bars are not made by the bar instability (see also Kormendy 2013). 

\begin{figure}[!tH]
\vbox{
\centerline{\psfig{figure=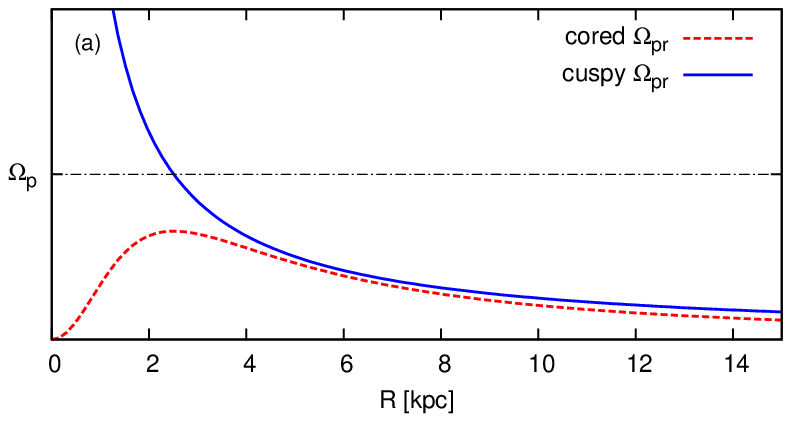,width=62mm,angle=0,clip=}\hspace{2mm} \psfig{figure=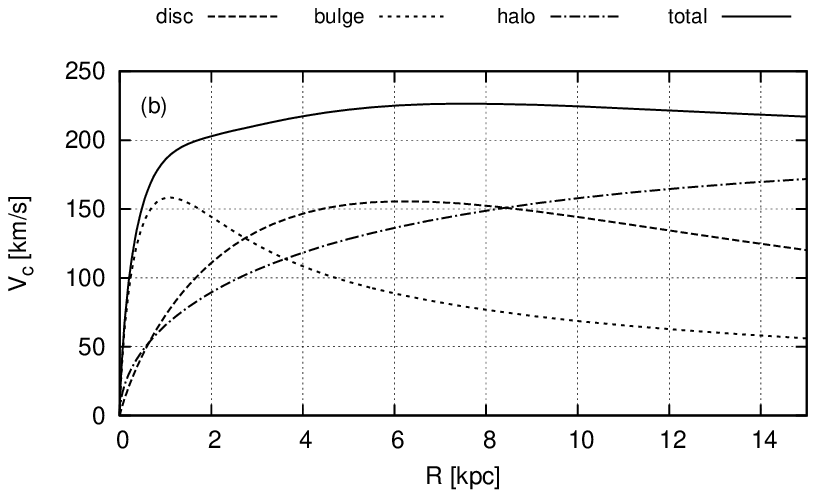,width=62mm,angle=0,clip=}}
\vspace{1mm}
\captionb{1}
{(a) The precession curves $\Omega_\textrm{pr}$ for typical cored (dashed line) and cuspy (solid line) galaxy models. (b) The total circular velocity profile $V_\textrm{c}(R)$ of the model with $z_\textrm{d} = 300$ pc and contributions of disc/bulge/halo components.}
}
\end{figure}

However, N-body simulations (Widrow et al. 2008 and references therein) show that usual bar instability is possible even in discs with cusps. We adopted the same galaxy model and recalculated the evolution with larger number of particles using different N-body schemes and varying key parameters such as velocity dispersion, disc thickness, mass of the disc. In all cases we see the bar forming sooner or later. 

The goal of this work is to explain a discrepancy between theory and N-body experiments and possibly explain the observations mentioned above. To achieve it, we analysed several possibilities, including possible insufficient bulge and disc particle resolution, lack of exact equilibrium, numerical accuracy of ILR determination, use of 3D instead of planar DF, gravity softening, dependence of the numerical scheme (tree code or particle-mesh), disc thickness. While all other reasons of the discrepancy but the last one were eliminated, it turned out that taking into account disc thickness allows to solve the problem even in the framework of linear theory.

\sectionb{2}{CALCULATIONS AND RESULTS}

We use a 3-component model by Widrow et al. (2008), consisting of the stellar disc, bulge, and dark matter halo. The disc is exponential, with radial scale $R_\textrm{d} = 2.9$\,kpc, truncation radius 15 kpc, mass $M_\textrm{d} =4.2\cdot 10^{10}$\,M$_\odot$. The radial velocity dispersion $\tilde \sigma_R$ is exponential, with central value $\sigma_{R0}=100$\,km/s and radial scale length $R_{\sigma} = 2R_\textrm{d}$. In the solar neighbourhood ($R=8$\,kpc), the radial velocity dispersion is $\sigma_{R}=25$\,km/s, the surface density is 50 M$_\odot$/pc$^2$. A characteristic height $z_\textrm{d}$, defined so that the surface density $\Sigma_d(R) = 2 z_d\rho(R, z=0)$, varies from 100 to 400 pc. 

Let assume a S{\'e}rsic bulge
\begin{equation}
\tilde \rho_\textrm{b}(r) = \rho_\textrm{b} \left( \frac r{R_e} \right)^{-p} \textrm{e}^{-b(r/R_e)^{1/n} }\ ,
\label{eq:bulge_dens}
\end{equation}
where $r$ is a spherical radius. For the S{\'e}rsic index $n=1.11788$, $R_e = 0.64$ kpc and adopted scale density, we have a bulge cusp index $p=p(n)\simeq 0.5$ (Golubov \& Just 2015), and mass $M_\textrm{b}=1.02\cdot 10^{10}$ M$_\odot$.

The target density profile of the halo is a truncated NFW profile with the scale $a_\textrm{h} = 17.25$\,kpc, truncation radius $r_\textrm{h} = 229.3$ kpc, and the total mass $M_\textrm{h}=1.29\cdot10^{12}$ M$_\odot$ (Diemand et al., 2008; Moetazedian \& Just 2016). Despite the halo density distribution is more cuspy than the bulge one, the latter dominates in the rotation curve down to $R \sim 0.01$ kpc, so the cusp index for the model is $\alpha = p \simeq 0.5$.  
 
Figure 1b shows the total circular velocity profile (solid curve) and contributions of separate components. The rotation curve is bulge-dominated at radii $R\lesssim 2.5$ kpc, and halo-dominated at $R > 9$ kpc. At radius $R\approx 6$ kpc, where the disc contribution peaks, the force from the halo is about 2/3 of the force from the disc in the galactic plane. Functions $\Omega(R)$ and $\kappa(R)$ diverges weakly at $R \to 0$ as $R^{-\alpha/2}$ with $\alpha \approx 0.5$. The Toomre $Q$ profile remains below 3 in the region $1 < R < 18$ kpc. The minimum $Q_\textrm{min} = 1.4$ is attained at $R= 5.9$ kpc. 

We performed 19 live runs, in which disc, bulge and halo are represented by particles, of the basic model: $z_\textrm{d} = 300$ pc, $M_\textrm{d}=4.2\cdot10^{10}$ M$_\odot$, $\sigma_{R0}=100$ km/s. The numerical simulations were carried out by the particle-mesh code Superbox-10 (Bien et al. 2013), and by the tree-code Bonsai-2 (B{\'e}dorf et al. 2012a,b). The total number of particles varied from N=5.6M to 104.5M. Our default runs have 16.75M, with 6M particles in the disc, 1.5M in the bulge, and 9.25M in the halo. Runs with smaller and higher number of particles are used to show the effect of a $N$-variation. The mass of halo particles in the largest simulation (104M in total) is only twice as heavy as the disc and bulge particle mass, so this run is used to show the absence of disc heating from heavier halo particles due to shot noise. 

Some runs denoted by `m' have multi-mass halo particles to achieve a better resolution in the bar region. In the region between 0.1 and 1 kpc, the number density ratio of our multimass and single mass runs varies from 10 to 100, thus the effective numerical resolution there is enhanced by this factor.

\begin{figure}[!tH]
\vbox{
\centerline{\psfig{figure=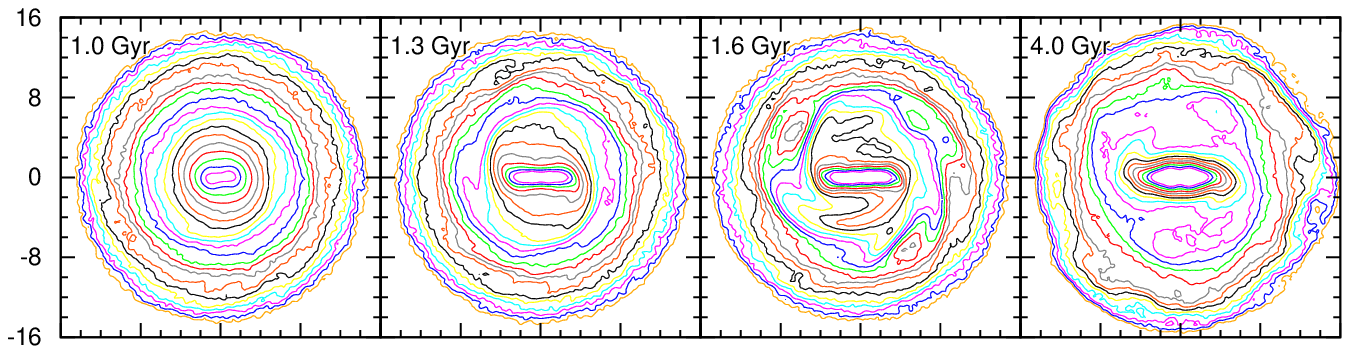,width=160mm,angle=0,clip=}}
\vspace{1mm}
\captionb{2}
{Bar patterns oriented along the x-axis on different stages of bar evolution. The curves are isolines of the density evenly spaced in log scale (10 levels for every factor of 10). Each frame size is 32x32 kpc.}
}
\end{figure}

Figure 2 shows bar isophotes for different stages of bar evolution. In this particular run the lag of bar formation is 700--800 Myr, so the first frame (1 Gyr) shows the bar in the beginning of the formation process. A lag in bar formation is often seen, especially when the number of disc particles is large, and growth rates of the instability is small. The second frame (1.3 Gyr) shows the bar at the moment of instability saturation; the bar radius is 4 kpc. However, it continues to increase until 1.6 Gyr and 4.9 kpc, then it begins to shorten. After instability saturation, the bar pattern speed decreases, and the corotation radius moves outwards. The last frame shows the bar at the end of the simulation.

The calculation of the components of the inertia ellipsoid is the simplest way to obtain the pattern speed and the growth rate of the bar mode. The former are obtained from an angle of the rotation of the main axes of the ellipsoid. Slopes of the bar strength $B(t)$ and the bar amplitude $A_2/A_0$, where
\begin{equation}
B(t) = 1 - I_{yy}/I_{xx}\ ,\quad A_m = \sum\limits_{j} m_j \textrm{e}^{-im\theta_j}\ ,
\label{eq:barstr}
\end{equation}
give very close estimates for the growth rate. Here $m_j$ and $\theta_j$ are mass and polar angle of star $j$; $j$ spans particles within some fixed radius (e.g., $R_d$), or within a growing domain that encompass the growing bar.

The results obtained in different runs are close to each other. For the pattern speed, the value estimated from Superbox runs is 51\,...\,52 km/s/kpc, the value estimated from Bonsai runs is 54\,...\,55 km/s/kpc. For the growth rates the value estimated from Superbox runs is 3.6\,...3.8 Gyr$^{-1}$, the value estimated from Bonsai runs is 4.2\,...4.4 Gyr$^{-1}$. For these pattern speeds, the ILR radius lies at $\simeq 0.4$ kpc. 

It is impossible to simulate the cuspy distribution perfectly with a finite number of particles. Besides, there are additional potential issues such as numerical accuracy, correctness of equilibrium, gravity softening, that can significantly affect the results of N-body simulations. Our estimate show however, that our models are well-resolved up to 0.1 kpc. Thus the discrepancy between theory and N-body experiment needs to be explained.   

To do it, we turn to the calculation of global modes using linear perturbation theory, in which special matrix equations are used (Polyachenko 2005, Polyachenko \& Just 2015). However, such equations are only available for razor thin discs, and they cannot take into account live bulge and halo. Thus, to compare correctly we need to perform rigid halo/bulge calculations, in which only the disc component is represented by particles. 

We made additional 6 rigid runs of the basic model, 12 rigid runs with different $z_\textrm{d}$, i.e. 18 rigid runs in total. The numerical simulations were carried out by the particle-mesh code Superbox-10, and by a self-coded Tree-GPU based gravity calculation routine {\tt ber-gal0}\footnote{\tt ftp://ftp.mao.kiev.ua/pub/users/berczik/ber-gal0/} 
(Zinchenko et al. 2015), which includes the expansion for force computation up to monopole order, with opening angle $\theta$ = 0.5. The number of disc particles varied from 1.1M to 6M.  

The pattern speeds $\Omega_\textrm{p}$ obtained in rigid runs are very close to the values of the live runs. This is in accordance with Polyachenko (2004) theory, in which global modes, such as bar modes, are density waves in a system of precessing orbits. The structure of orbits is independent on the nature of bulge and halo components, thus, the pattern speeds should be the same.

In contrast, the growth rates $\omega_\textrm{I}$ of the rigid runs are significantly lower (by a factor of 2) than $\omega_\textrm{I}$ in live runs. Considerably weaker bars in the rigid halos were also obtained by Athanassoula (2002), and explained by the additional interaction of the bar with halo particles, mainly on the corotation resonance.  

A new effect that we noticed, especially in rigid runs, is some uncertainty of the N-body results manifested in different lags and values of growth rates. So, $\omega_\textrm{I}$ found in rigid runs was in the range between 1.1 to 1.9 Gyr$^{-1}$. This effect is seen even in our large tree-code simulations with $N_\textrm{d}=6$\,M in which we use small fixed gravity softening equal to 10 pc. This effect is possibly related to stochasticity effects noted by Sellwood \& Debattista (2009).

Ignoring the difference between thick and razor-thin discs, one can try to reproduce unstable bar modes of the rigid models by the matrix method. As a `zero-order' approximation, one can integrate a disc density over $z$-component. However, no unstable modes were found. 

The `first-order' approximation is to take into account the elevation of particles above the disc plane and calculate an effective radial force. The force exerted on the particle at height $z$, $F_R = -\p\Phi(R,z)/\p R$, is smooth at $R=0$ and $z\ne 0$, and provides finite $\Omega(R,z)$ and $\kappa(R,z)$. Thin lines in the left panel of Figure 3 show $\Omega_\textrm{pr}(R,z)$ for different elevations above the plane. The thick solid line shows an mass weighted average precession curve. It is crucial that it has a maximum, in contrast to the cuspy precession curve measured in the equatorial plane (dashed curve). All frequencies $\Omega_\textrm{p}$  above the maximum (43.7 km/s/kpc) are available as pattern speed for bar formation.  
  
\begin{figure}[!tH]
\vbox{
\centerline{\psfig{figure=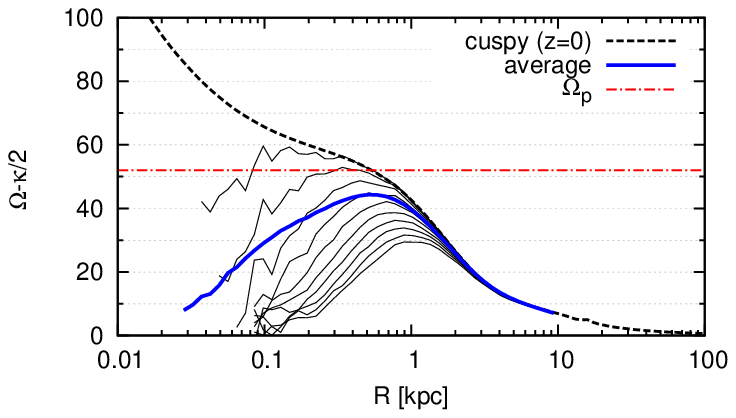,width=62mm,angle=0,clip=}\psfig{figure=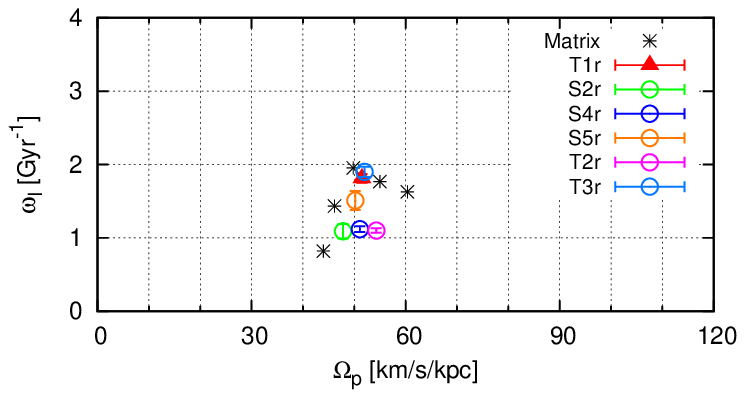,width=62mm,angle=0,clip=}}
\vspace{1mm}
\captionb{3}
{The basic model ($z_\textrm{d}=300$ pc). {\it Left panel:} The precession curves for particles with different elevation above the plane (each thin solid curve is calculated for particles in the layer $\Delta z = 50$ pc width, starting from $z=0$), and average $\overline \Omega_\textrm{pr}$ (thick solid curve). The cuspy profile in the equatorial plane is shown by the dashed line. The dash-dotted line shows the bar pattern speed obtained in rigid halo/bulge simulations. {\it Right panel:} Eigenmodes obtained with the matrix method are show by black asterisk, other signs show N-body eigenmodes (prefix `S' is for Superbox, `T' -- for tree-code). }
}
\end{figure}

A  comparison of the N-body modes determined from the rigid runs and results of eigenmodes calculation using the matrix equation by Polyachenko (2005) is given on the right panel of Figure 3. All unstable matrix modes avoid a ILR. The pattern speeds of N-body modes are close to the average pattern speeds of the matrix modes. The growth rates of N-body modes suffer from stochasticity, but are also in agreement with the matrix calculations.

Using a set of 12 rigid runs we followed the dependence of the pattern speeds and growth rates vs. disc thickness $z_\textrm{d}$. The results are given in Figure 4. For the pattern speeds, agreement is good for all vertical scale length in the range 100 ... 400 pc. The growth rates have an outlier at $z_\textrm{d} =100$ pc; the reason is not yet clear. The N-body modes fit well the matrix predictions in the range 150...300 pc, however then the growth rate fall of the smooth fit for the matrix values. The plausible reason for the discrepancy seen in relatively thick models is the increasing vertical velocity dispersion that stabilises the disc in the same way as the radial velocity dispersion.
 
\begin{figure}[!tH]
\vbox{
\centerline{\psfig{figure=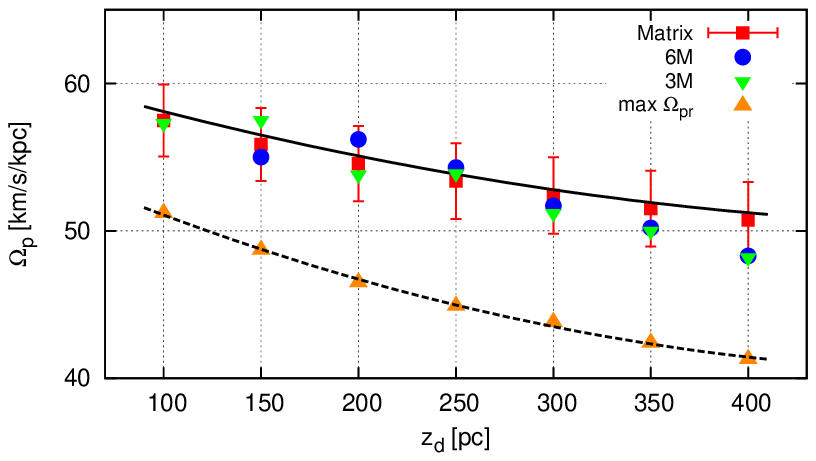,width=62mm,angle=0,clip=}\psfig{figure=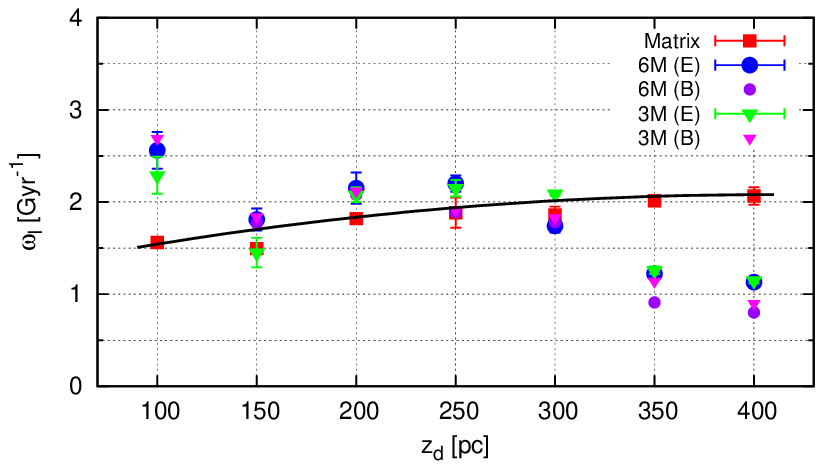,width=62mm,angle=0,clip=}}
\vspace{1mm}
\captionb{4}
{Comparison of characteristics of eigenmodes for different disc vertical scale length $z_\textrm{d}$ obtained in the matrix method and N-body $N_\textrm{d}=6$M and 3M runs. {\it Left panel:} The pattern speeds. Upper and lower limits of the red error bars show pattern speeds of modes with the largest growth rates. Orange upward triangles show maxima of the precession curves. Black solid and dashed lines are smooth fits.  {\it Right panel:} The growth rates. Upper and lower limits of the red error bars show two maximum growth rates.}
}
\end{figure}

Matrix calculations show a strong dependence of the growth rates on the disc mass. For example, for $M_\textrm{d}$ only 11 per cent lower than adopted in our models, the obtained growth rates are 2.5 times smaller (Figure 5, left panel). If one can extrapolate these results to live discs preserving the ratio of the growth rates of the live and rigid models at factor of 2, it means that the galactic disc remains nearly stable (instability time is larger than age of the Universe) for a long time, and the bar went through formation recently. 

\begin{figure}[!tH]
\vbox{
\centerline{\psfig{figure=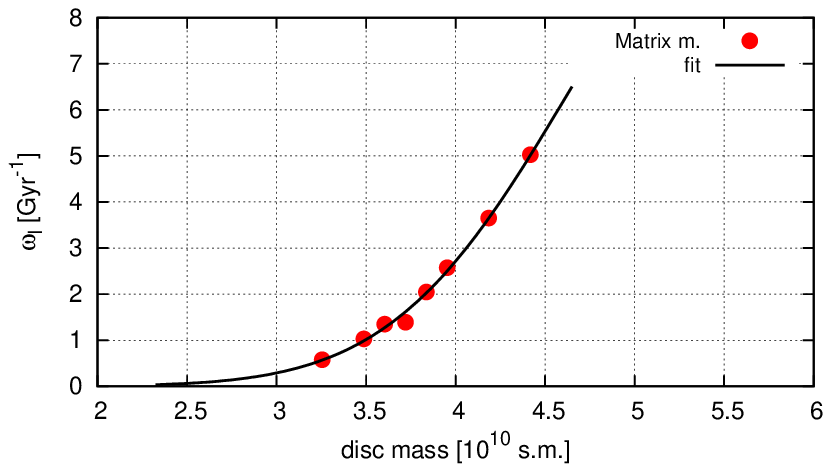,width=62mm,angle=0,clip=}\psfig{figure=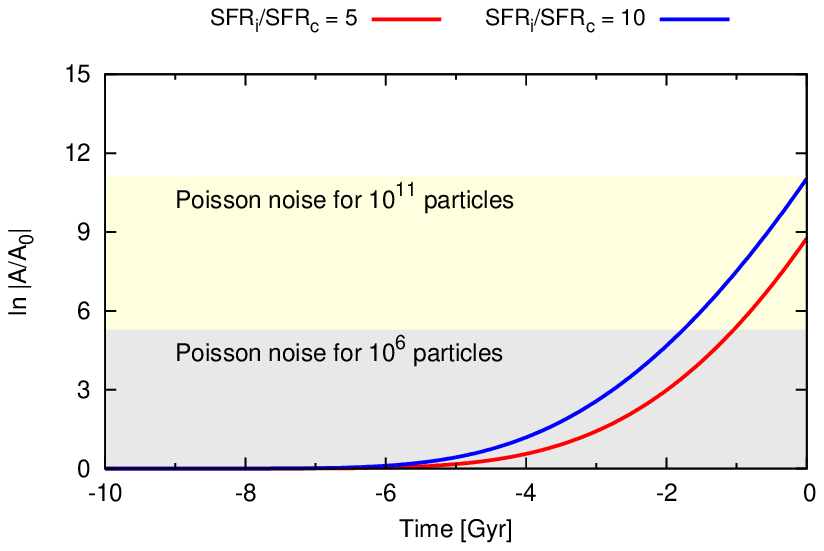,width=62mm,angle=0,clip=}}
\vspace{1mm}
\captionb{5}
{{\it Left panel:} The growth rate estimates vs. disc mass $M_\textrm{d}$. {\it Right panel:} Amplification of the initial perturbation taking into account growth of the disc mass according to adopted SFR profiles (see main text).}
}
\end{figure}

Our crude estimates using exponential star formation rates (Aumer \& Binney 2009), with initial to current ratio equal to 5 and 10, shows that during 10 Gyr small disc perturbations grow to factor $\exp(9...11)$ (Figure 5, right panel). To compare, factor $\exp(6)$ is needed for the Poisson noise of disc consisting of $N_\textrm{d} \sim 10^6$ particles to grow into a bar; factor $\exp(11)$ is needed for the noise in $N_\textrm{d} \sim 10^{11}$ particles. Certainly the Poisson noise level for $10^{11}$ particles is unreasonably low because of the presence of giant molecular clouds. A reasonable estimate is that bar have undergone the instability saturation 1 ... 2 Gyr ago.

\sectionb{3}{CONCLUSIONS}

In this paper we study bar formation in a galactic model with a cuspy bulge, exponential disc, and NFW halo. Using accurate $N$-body simulations with particle--mesh (PM) and Tree codes, with more than 100M particles, we have shown without doubts that a bar is formed despite the presence of the ILR. We argue that this discrepancy can be explained by taking into account the finite disc thickness.

Using a simple model, in which the radial force is averaged over the $z$-coordinate perpendicular to the disc plane, in global mode analysis of linear perturbation theory we obtained pattern speeds and growth rates of unstable modes that agree well with N-body, for different vertical scale length of the disc. 

The substitution of halo and bulge particles by a rigid external potential revealed that rigid cuspy models are less unstable than live ones, i.e. typical growth rates are a factor of two smaller. This is in agreement with results by Athanassoula (2002) obtained for bulgeless models. At the same time, pattern speeds in live and rigid runs are close (relative difference is 5 per cent or less).

The stochastic behaviour of N-body models, mentioned e.g. by Sellwood and Debattista (2009) for disc evolution after bar formation, is seen in our runs also, especially in case of the rigid halo and bulge, when growth rates are small. This manifests itself in the appearance of a random lag before the exponential growth of the amplitude, and sometimes in a non-exponential character of the growth. 

For the usual bar mode instability in thin discs, the behaviour of the $\Omega_\textrm{pr}$ profile determining the position of the ILR is important. However, particles that elevate above the equatorial plane do not feel central angular velocity singularity, and radial force averaged over vertical axis provides cored, rather than cuspy, $\Omega_\textrm{pr}$ profile. This means that ILR is practically non-existent at $R \lesssim z_\textrm{d}$.

The strong dependence of the growth rates on disc mass favours the hypothesis of recent bar formation. This fact can explain the observed lack of barred galaxies at redshifts $z\gtrsim0.5$, and low ratios of corotation to bar radii, $0.9 < {\cal R} < 1.3$.

\thanks{The main production runs was done on the {\tt MilkyWay} supercomputer, funded by the Deutsche
Forschungsgemeinschaft (DFG) through the Collaborative Research Centre (SFB 881) ``The Milky Way System'' 
(subproject Z2), hosted and co-funded by the J\"ulich Supercomputing Center (JSC). The special GPU 
accelerated supercomputer {\tt laohu} at the Center of Information and Computing at National 
Astronomical Observatories, Chinese Academy of Sciences, funded by Ministry of Finance of People's 
Republic of China under the grant $ZDYZ2008-2$, has been used for some of code development. 
We also used smaller GPU cluster  {\tt kepler}, funded under the grants I/80 041-043 and I/81 396 of 
the Volkswagen Foundation and grants 823.219-439/30 and /36 of the Ministry of Science, Research and 
the Arts of Baden-W\"urttemberg, Germany. 
This work was supported by the Sonderforschungsbereich SFB 881 ``The Milky Way System'' (subproject A6) 
of the German Research Foundation (DFG), and by the Volkswagen Foundation under the Trilateral Partnerships 
grant No. 90411. E.P. acknowledges a financial support by Russian Basic Research Foundation, 
grants 15-52-12387, 16-02-00649, and by Basic Research Program OFN-15 `The active processes in galactic 
and extragalactic objects' of Department of Physical Sciences of  RAS. P.B. acknowledges the special 
support by the NASU under the Main Astronomical Observatory GRID/GPU {\tt golowood} computing cluster 
project.}

\References

\refb Abraham R. G. et al. 1999, MNRAS, 308, 569

\refb Athanassoula E. 2002, ApJ, 569, 83

\refb {B{\'e}dorf} J., {Gaburov} E., {Portegies Zwart} S. 2012a, in
  Astronomical Society of the Pacific Conference Series, Vol. 453, Advances in
  Computational Astrophysics: Methods, Tools, and Outcome, {Capuzzo-Dolcetta}
  R., {Limongi} M., {Tornamb{\`e}} A., eds., p. 325

\refb {B{\'e}dorf} J., {Gaburov} E., {Portegies Zwart} S. 2012b,
  {Bonsai: N-body GPU tree-code}. Astrophysics Source Code Library

\refb {Bien} R., {Just} A., {Berczik} P., {Berentzen} I. 2008, Astronomische Nachrichten, 329, 1029

\refb J. Binney, S. Tremaine,{\it Galactic Dynamics} (Princeton : Princeton University Press, 2008)


\refb Diemand J. et al. 2008, Nature, 454, 735


\refb Golubov O., Just A. 2013, IAUS, 295, 231 

\refb Kormendy J. {\it Secular Evolution of Galaxies} (Cambridge : Cambridge University Press, 2013)

\refb Mark J. W.-K. 1971, Proc. Natl. Acad. Sci. USA, 68, 2095 

\refb Mark J. W.-K. 1974, ApJ, 193, 539

\refb Merrifield M. R., Abraham R. G., Ellis R. S., Tanvir N. R., Brinchmann J. 2000, ASP Conf. Ser., 197, 39

\refb Moetazedian R., Just A. 2016, MNRAS, 459, 2905

\refb Polyachenko E. V. 2004, MNRAS, 348, 345

\refb Polyachenko E. V. 2005, MNRAS, 357, 559

\refb Polyachenko E. V., Just A. 2015, MNRAS, 446, 1203

\refb Sellwood J. 2000, ASP Conf. Ser., 197, 2

\refb Sellwood J.A., Debattista V.P. 2009, MNRAS, 398, 1279

\refb Toomre A., {\it Structure and evolution of normal galaxies} (Ed. S.M. Fall and D. Lynden-Bell, Cambridge Univ. Press., 1981), p.111.

\refb Widrow L. M., Pym B., Dubinski J. 2008, ApJ, 679, 1239

\refb {Zinchenko}, I.~A., {Berczik}, P., {Grebel}, E.~K., {Pilyugin}, L.~S., {Just}, A. 2015, ApJ 806, 267

\end{document}